\documentclass[letterpaper,10pt]{article}
\usepackage{osameet2}

\usepackage{amsmath,amssymb}

\usepackage{textcomp}

\begin{document}
	
	\vspace{-2mm}

	\title{White Gaussian Noise Based Capacity Estimate and Characterization of Fiber-Optic Links}
	\vspace{-6mm}
	
	\author{
		Roland~Ryf\textsuperscript{(1)},
		John van Weerdenburg\textsuperscript{(2)},
		Roberto A. Alvarez-Aguirre\textsuperscript{(1,3)},
		Nicolas~K.~Fontaine\textsuperscript{(1)},
		Ren\'{e}-Jean Essiambre\textsuperscript{(1)},
		Haoshuo~Chen\textsuperscript{(1)},
		Juan Carlos Alvarado-Zacarias\textsuperscript{(1,3)},		
		Rodrigo~Amezcua-Correa\textsuperscript{(3)},
		Ton Koonen\textsuperscript{(2)},
		Chigo Okonkwo\textsuperscript{(2)}
	}
	
	\maketitle                  
	
	
	\address{
		
		\textsuperscript{(1)} Nokia Bell Labs, 791 Holmdel Rd, Holmdel, NJ, 07733, USA
		
		\textsuperscript{(2)} Institute for Photonic Integration, Eindhoven University of Technology, 5600 MB Eindhoven, Netherlands
		
		\textsuperscript{(3)} CREOL, The University of Central Florida, Orlando, Florida 32816, USA
	}
	\email{ Roland.Ryf@nokia.com}
	\vspace{-4mm}
	\begin{abstract}
		We use white Gaussian noise as a test signal for single-mode and multimode transmission links and estimate the link capacity based on a calculation  of mutual information. We also extract the complex amplitude channel estimations and mode-dependent loss with high accuracy.
	\end{abstract}
	\vspace{0mm}

\vspace{-4mm}
\section{Introduction}
\label{sec:intro}
\vspace{-1mm}

White Gaussian noise (WGN) is the capacity maximizing signal according to the Shannon theory for a channel with additive white Gaussian noise (AWGN), and in practice the formats used in the latest digital coherent transmission experiments are slowly converging to white Gaussian noise if we consider pulse shaping, Nyquist WDM, higher order QAM formats, and most recently probabilistic constellation shaping.
Which begs the question: Why not use white Gaussian noise as test signal for fiber transmission systems?
The are some fundamental advantages in doing so: The WGN signal has an infinite pattern length, which eliminates many of the common decorrelation issues between WDM and also SDM channels encountered when working with periodic test signals. Also, a WGN signal is symbol-rate agnostic, and can be interpreted as any desired baud-rate, or even as an OFDM signal as long as the signal bandwidth  fits within the receiver bandwidth, and is therefore testing the fiber channel in a  generic way.
The difficulty is, however, that the WGN signals never repeats and therefore must be captured not only after transmission like in conventional transmission experiments, but also before transmission. This can however be easily achieved by using the  sequence trigger functionality that is available in most digital storage oscilloscopes (DSO). 

In this work, we demonstrate how a WGN signals can be used to estimate the capacity of a transmission link by calculating the mutual information as function of the length and launch power into a single-mode fiber (SMF) and a few-mode fiber (FMF) with 3 spatial modes. Additionally, we also show exemplary results of channel estimations with more than 60~dB dynamic range and mode-dependent loss (MDL) calculations that are also obtained by using WGN test signals.

White Gaussian noise based transmission measurements are not going to replace traditional high performance transmitter based experiments, however they provide a simple, reproducible and generic capacity estimate and a solid characterization technique for fiber-optic links that can be realized with a minimal amount of test equipment.

\vspace{-0.2cm}
\section{WGN based transmission experiment}

\begin{figure}[bthp]
	\vspace{-0.3cm}
	\centering
	\includegraphics[width=0.8\columnwidth]{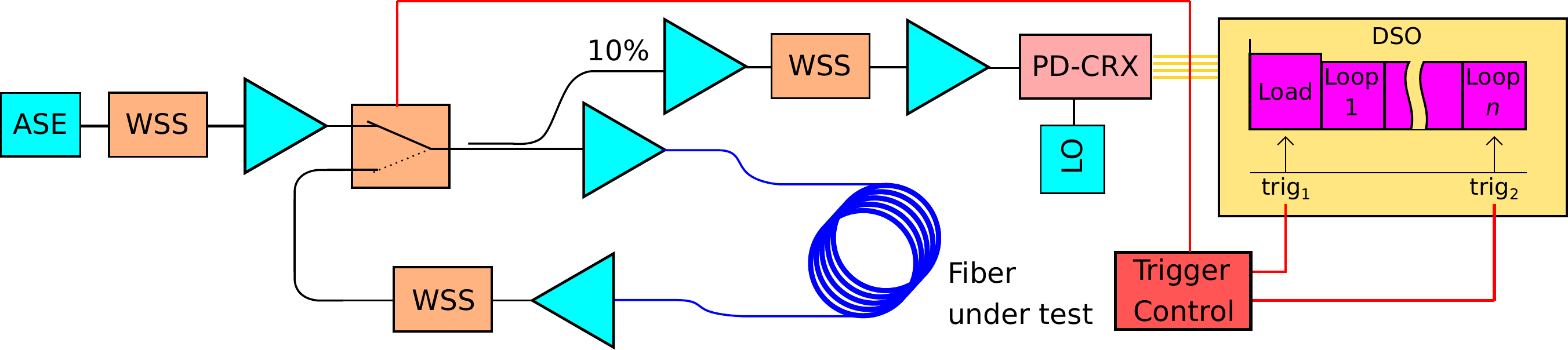}
	\vspace{-0.3cm}
	\caption{Setup for WGN transmission experiment}
	\label{fig:setup}
	\vspace{-0.4cm}
\end{figure}
Gaussian noise is widely used as light source for various optical characterization technique like spectral transmission measurements, low-coherence interferometry, or crosstalk measurements. 
WGN has also been used in transmission system characterization for example to emulated nonlinear crosstalk effect by replacing neighboring wavelength channels with noise\cite{Elson:16}, or to estimate nonlinear noise in few-mode fibers\cite{Ellis:13}.
In this work, we go one step further by using WGN for the signal under test. Because WGN is a truly random signal, this is only possible by capturing a copy of the same noise instantiation before and after transmission through the fiber under test. 
The basic experimental setup for the measurement is shown in Fig.~\ref{fig:setup}. 
We use amplified spontaneous emission (ASE) of an erbium-doped amplifier (EDFA) to generate Gaussian noise. The generated noise is then spectrally equalized by using a wavelength selective switch (WSS). The resulting WGN signal is 
than captured before and after transmission through the fiber under test by using a polarization diverse coherent receiver (PD-CRX).
As local oscillator (LO) for the coherent receiver, we used three different  narrow linewidth lasers with linewidth ranging from sub KHz to sub Hz (Pure-Photonics ITLA PPCL100 $<10$~KHz, Rio Orion $< 1$~KHz,  OEwaves $< 1$~Hz), where the latter offers phase stabilities over millions of samples, which is helpful to simplify the phase recovery of the transmitted signal.

The method naturally fits with recirculating loop experiments, where the WGN signal can be captured when injected into the loop and the second time after the desired number of recirculations. This can be achieved by using the sequence trigger function of a digital sampling oscilloscope, which allows for multiple triggers and related capture windows within a single data acquisition.
The loop consists of a solid state 1x2 switch to switch between loading and recirculation, a 10/90 splitter to extract the loop signal, a two-stage amplifier and a WSS to precisely equalize the spectrum in the loop. The light extracted from the loop is amplified by a second two-stage EDFA, where a WSS is placed between the stages to select the desired spectral region to be captured.
The captured data consists of two time windows including 8 Million samples each captured at 40~GS/s.
The method can be extended to support mode-multiplexed transmission experiments by replacing all components with multimode components or by using parallel loops, in which case either a single WGN signal can be split and delay decorrelated before coupling into the individual modes as commonly done in mode-multiplexing experiments or multiple independent WGN sources can be used as they are readily available. In our experiment we performed transmission using a 400~GHz wide WGN channel and we used two fibers under test, the first was a 78~km long standard single-mode fiber, and the second was a depressed cladding large-effective area 3-mode fiber with a length of 96~km\cite{Ryf2012}. 
The methods also allows for WDM-type measurements, that can be performed by tuning the wavelength of the LO in steps comparable to the bandwidth of the receiver, therefore covering the whole bandwidth of the generated WGN signal. 
\begin{figure*}[!th]
	\centering
	\includegraphics[width=1.0\columnwidth]{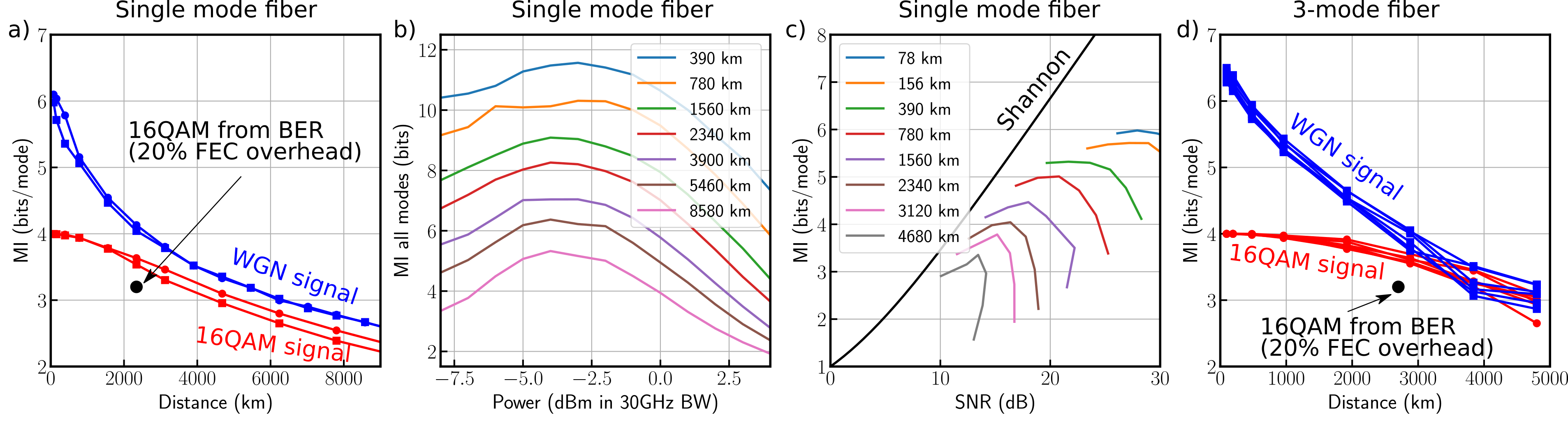}
	\vspace{-0.6cm}
	\caption{a) Mutual information as function of distance for a 78-km SMF  measured with a WGN signal and 16~QAM signal, and b) plotted as function of launch power  and c) as function of measured SNR. d) Mutual information as function of distance for a 96~km long 3-mode fiber.}
	\label{fig:smfmidist}
	\vspace{-0.6cm}
\end{figure*}

\section{Digital signal processing for WGN signals}
As WGN signals do not contain any inherent structure, many traditional blind algorithms commonly used for clock-recovery, frequency offset and phase estimation cannot be directly applied. 
This can be overcome by using data-aided methods, which in practice means that training sequences are required, and in particular for phase tracking, it is useful to have a narrow  bandwidth laser as local oscillator, so that training sequences can be used less frequently.
In this work the data are processed in the following manner: The captured data are upsampled from 40GS/s to 60GS/s, filtered with 15-GHz high-order Gaussian filter, and electronic dispersion compensation according to the fiber length is applied to the transmitted signal, after which the data is treated as a two times oversampled 30~Gbaud signal, with continuous levels.
Note that choice of the baud rate is arbitrary, and the same measurement can be evaluated considering different baud rates, as long as the resulting bandwidth is smaller than the receiver bandwidth. 
We define the input field $F_{{\textrm in},m}(n)$, where $m$ is the index identifying either the polarization or the mode and $n$  ranges from 1 to $Ns$, where $Ns$ is the number of acquired samples. Similarly to $F_{{\textrm in},m}(n)$ we define the output field $F_{{\textrm out},m}(n)$  and subsequently we evaluate the cross-correlation between the fields in order to align them in time. Note that the LO with $< 1$~Hz linewidth will produce high dynamic range correlations peaks as the relative phase between the fields is stable over millions of samples and the signals has a infinite pattern length. In practice however, tunable ITLA can be used for the measurement. 
\begin{figure}[!t]
	\vspace{-0.4cm}
	\centering
	\includegraphics[width=1.0\columnwidth]{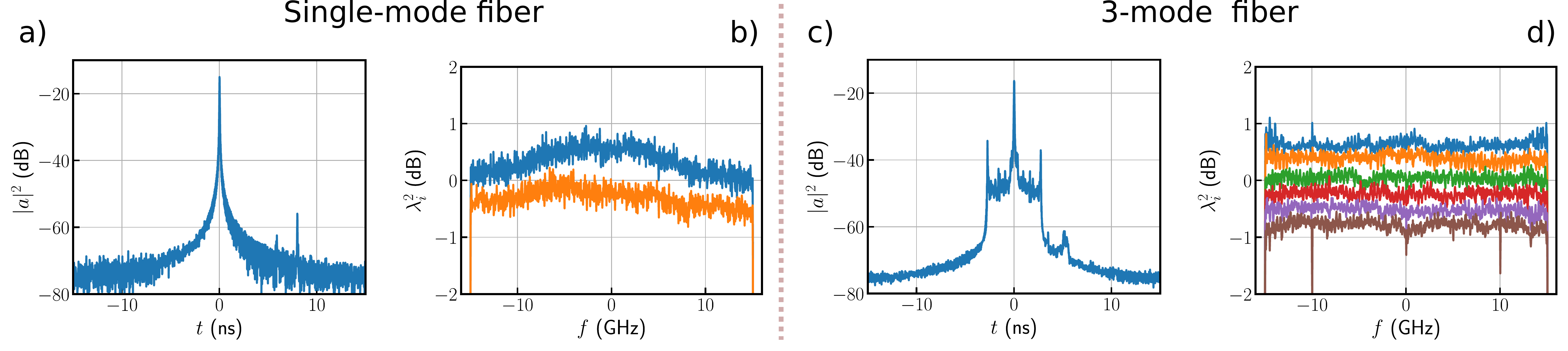}
	\vspace{-0.6cm}
	\caption{Example of impulse responses and MDL WGN based measurements: a) Impulse response and b) PDL of a SMF fiber after 1560-km transmission, whereas  c) shows the impulse response and d) MDL after 96~km 3-mode fiber. }
	\vspace{-0.6cm}
	\label{fig:imprespexamples}
\end{figure}
The fields are than processed by a frequency domain equalizer (FDE) that is optimized according to the data-aided least mean square (LMS) algorithm, where $F_{in}$ is used as data and $F_{out}$ is processed by FDE and producing the equalized field  $F_{{\textrm eq},m}(n)$.
Subsequently, a data-aided phase recovery with an averaging window of 200 samples is used to verify that the phase differences between  $F_{{\textrm in},m}$ and $F_{{\textrm eq},m}$.
In order to estimate the capacity of the transmitted signal we calculated the mutual information (MI) between $F_{{\textrm in},m}$ and $F_{{\textrm eq},m}$, which is calculated based on multi ring constellations\cite{Goebel2011}, where we used 16  optimally spaced rings that make use of  the radial symmetry of the WGN signal. 
The results are plotted in Fig.~\ref{fig:smfmidist}a for a 78-km standard single-mode fiber, where the mutual information for each polarization is plotted separately. The MI for both polarizations is shown as function of the launch power in Fig.~\ref{fig:smfmidist}b, where the power is normalized to a 30~GHz reference bandwidth. Further, the results are plotted in Fig.~\ref{fig:smfmidist}c as function of the received SNR measured with an optical spectrum analyzer. Finally, Fig.~\ref{fig:smfmidist}d shows the MI for all 6 spatial tributary of a 3-mode fiber, showing that the method can also be used to estimate the capacity of mode-multiplexed systems.
To confirm the results we also performed conventional transmission using 12 WDM channels spaced at 33.33~GHz and modulated at 30-Gbaud with a 16~QAM polarization multiplexed signal. The extracted mutual information is also plotted in  Fig.~\ref{fig:smfmidist}a and Fig.~\ref{fig:smfmidist}d as reference, confirming the validity of the proposed method, which achieves slightly  higher MI than the modulated signal at longer distance and is significantly higher at shorter distances where 16~QAM is limited to 4 bits/mode.
The method can also be used to perform a channel estimation. This can be achieved by inverting the roles of $F_{{\textrm in},m}$ and  $F_{{\textrm out},m}$ in the FDE equalizer. After convergence of the FDE, the equalizer weights will contain the channel estimation in the frequency domain. The channel matrix can subsequently be analyzed by performing a singular value decomposition, which will provide the mode-dependent loss (MDL) as function of the frequency. As an example, the polarization dependent loss (PDL) of a SMF is shown in Fig.~\ref{fig:imprespexamples}b whereas Fig.~\ref{fig:imprespexamples}d shows the MDL of a 96~km long 3-mode fiber. 
The impulse response
of the channel is obtained by a Fourier transform of the equalizer coefficients
and the impulse response of a SMF after 1560 km
and a 3-mode fiber after 96 km are shown in Fig.~\ref{fig:imprespexamples}c and Fig.~\ref{fig:imprespexamples}d, respectively,
 demonstrating the large
dynamic range that the proposed method offers.

In conclusion we show that white Gaussian noise can be advantageously used as test signal for optical communication link. The proposed method is accurate and provides a full channel characterization and capacity estimate requiring a minimal amount of test equipment.

We acknowledge the Dutch NWO Graduate Photonics program and 
thank OFS for providing the 3-mode fiber and S.~Chandrasekhar for support with the narrow bandwidth lasers.

\vspace{-2mm}

\bibliographystyle{IEEEtran}


\end{document}